\documentclass[aps,physrev,twocolumn,superscriptaddress,showkeys,nofootinbib,10pt]{revtex4-2}
\usepackage{graphicx}
\usepackage{graphics}
\usepackage{amsmath}
\usepackage{amssymb}
\usepackage{amsfonts}
\usepackage{cancel}
\usepackage{dsfont}
\usepackage{braket}
\usepackage{color}
\usepackage{braket,slashed}
\usepackage[mathscr]{euscript}
\definecolor{darkblue}{rgb}{0, 0, 0.8}
\usepackage[dvipsnames]{xcolor}
\usepackage[colorlinks,bookmarks=false,citecolor=RoyalBlue,linkcolor=BrickRed,urlcolor=OliveGreen]{hyperref}
\usepackage{subfigure}
\usepackage{xfrac}
\usepackage{bm}
\usepackage{kantlipsum}
\usepackage{enumitem}
\usepackage{tikz}
\usepackage{framed}
\usepackage{graphicx}
\usepackage{subfigure}
\usepackage{cleveref}
\usepackage{array}

\usepackage[linesnumbered,ruled,vlined]{algorithm2e}
\usepackage{algpseudocode}
\usepackage{algorithmicx}
\usepackage{algcompatible}
\allowdisplaybreaks[1]

\newcommand{\code}[1]{\texttt{#1}}

\newcommand{\one}{\mathds{1}}

\newcommand{\e}{\ensuremath{\mathrm{e}}}

\begin{document}

\title{Matrix Product Operator Encodings of the Magnus Expansion and Dyson Series}

\author{Victor Vanthilt}
\thanks{Contact author: \href{mailto:victor.vanthilt@ugent.be}{victor.vanthilt@ugent.be}}
\affiliation{Department of Physics and Astronomy, Ghent University, Krijgslaan 299, 9000 Gent, Belgium}

\author{Maarten Van Damme}
\affiliation{SandboxAQ, 780 High Street, Palo Alto, California 94301, United States}

\author{Jutho Haegeman}
\affiliation{Department of Physics and Astronomy, Ghent University, Krijgslaan 299, 9000 Gent, Belgium}

\author{Ian P. McCulloch}
\affiliation{Department of Physics, National Tsing Hua University, Hsinchu 30013, Taiwan}
\affiliation{Frontier Center for Theory and Computation, National Tsing Hua University, Hsinchu 30013, Taiwan}

\author{Laurens Vanderstraeten}
\affiliation{Center for Nonlinear Phenomena and Complex Systems, Universit\'e Libre de Bruxelles, CP 231, Campus Plaine, 1050 Brussels, Belgium}

\date{\today}

\begin{abstract}
We introduce a matrix product operator (MPO) encoding of the Magnus expansion and the Dyson series for one-dimensional quantum lattice models with time-dependent Hamiltonians. The MPO construction can be made accurate up to arbitrary order in the time step, it can be applied to both finite and infinite systems, and it can handle long-range interactions. The resulting MPO can be combined with state-of-the-art time evolution algorithms based on matrix product states, allowing for drastic improvements in simulating evolution under time-dependent Hamiltonians. Our MPO construction can also be used for the optimization of quantum circuits in the context of quantum simulation of time-dependent Hamiltonians.
\end{abstract}

\maketitle

\section{Introduction\label{sec:Introduction}}

One of the most fundamental questions in modern quantum many-body physics revolves around integrating the time-dependent Schrödinger equation
\begin{equation}
    \partial_t \ket{\Psi(t)} = -i H(t) \ket{\Psi(t)},
\end{equation}
where $H(t)$ is, in general, a time-dependent Hamiltonian. For a given initial state at time $t_0$, the solution of this differential equation is formally written as
\begin{equation}
    \ket{\Psi(t)} = U(t,t_0) \ket{\Psi(t_0)}
\end{equation}
in terms of the time-evolution operator $U(t,t_0)$, a solution of the differential equation
\begin{equation}
\label{eq:schrodinger-timeevooperator-equation}
\partial_t U(t,t_0) = - iH(t)U(t,t_0), \qquad U(t_0,t_0)=\one \;.
\end{equation}
When the Hamiltonian is time independent, the solution for the time-evolution operator is simply given by the exponential $U(t, t_0) = \exp(- i H (t-t_0))$.  In contrast, for time-dependent Hamiltonians where the commutator $\left[H(t), H(t')\right] \neq 0$, the solution is formally denoted in terms of a time-ordered exponential
\begin{equation} \label{eq:time-ordered-exponential}
    U(t,t_0) = \mathcal{T}\exp \left( -i\int_{t_0}^{t} H(t')dt' \right) \;.
\end{equation}
However, this formal expression does not generally admit a closed form analytical evaluation, even for the simplest quantum systems, such as a single qubit. The value of $U(t, t_0)$ thus needs to be approximated using numerical integration of the differential equation, or via one of a number of analytical expansions that make the structure of the time-ordered exponential explicit.

A first option is to make the time-ordered exponential operational by dividing the time evolution into $M$ steps
\begin{equation}
    U(t,t_0) = \prod_{i=0}^{M-1} U(t_{i+1},t_i), \quad t_i=t_0+\frac{i(t-t_0)}{M}
\end{equation}
where one hopes that efficient ways can be found for approximating the time-evolution operator in each small time segment, e.g.~by considering the Hamiltonian effectively constant within the intervals $[t_i, t_{i+1}]$.

A second option is obtained by recursively integrating Eq.~\eqref{eq:schrodinger-timeevooperator-equation}, which gives rise to a uniformly converging series known as the Dyson series \cite{Dyson1949,Feynman1951}:
\begin{multline} \label{eq:dyson-series}
U(t, t_0) = \sum_{n=0}^{+\infty} (-i)^n \int_{t_0}^{t}dt_1 \int_{t_0}^{t_1}dt_2 \;\dots \int_{t_0}^{t_{n-1}}dt_n \\ \left[H(t_1)\dots H(t_n)\right],
\end{multline}
which reduces to the Taylor series of the exponential in the case of a time-independent Hamiltonian. The time ordering in the exponential is reflected in the different integration regions such that the operator product in the integrand is, in fact, time ordered. The Dyson series is an expansion in the total time interval $\Delta t=t-t_0$, so we expect it to converge more quickly as we decrease the latter.

A third approach is given by the Magnus expansion \cite{Magnus1954, Blanes2009}, which expresses the time-evolution operator as the exponential of an effective time-independent operator $U(t,t_0) = \e^{\Omega(t,t_0)}$, where $\Omega(t,t_0)$ is expressed as an expansion of the form
\begin{equation} \label{eq:magnus}
    \Omega(t,t_0) = \sum_n \Omega_n(t,t_0)
\end{equation}
with the first two terms given by
\begin{equation} \label{eq:magnus_operators}
\begin{aligned}
    &\Omega_1(t,t_0) = -i\int_{t_0}^t dt' H(t') \\
    &\Omega_2(t,t_0) = -\frac{1}{2!} \int_{t_0}^t dt_1' \int_{t_0}^{t_1'} dt_2' \; [H(t_1'),H(t_2')] \;.
\end{aligned}
\end{equation}
When the Hamiltonian is time-independent, the only non-zero term is \(\Omega_1(t)\), which reduces to the familiar value \(-iH(t-t_0)\). The first-order Magnus expansion thus treats the system as if it were evolving under a constant Hamiltonian that is the average over the time interval. The higher-order terms are corrections hereto.

For interacting quantum many-body systems, the exponentially large Hilbert space complicates each of these strategies. Even the normal exponential of a time-independent Hamiltonian cannot be efficiently computed or applied to a given initial state without approximation strategies. The most accurate methods for simulating the real-time evolution of quantum many-body systems on classical computers are provided by the formalism of tensor networks \cite{Schollwoeck2011, Cirac2021, xiang2024}. For one-dimensional and quasi-one-dimensional systems, the class of matrix product states (MPS) provides powerful simulation methods that efficiently capture low-energy states of local Hamiltonians; in this context, a plethora of performant time evolution methods have been developed \cite{Paeckel2019}; however, the question of how to approximate long simulation times efficiently remains a challenging open problem that continues to receive widespread attention. In the early days of the density matrix renormalization group (DMRG), the Trotter-Suzuki decomposition \cite{Trotter1959, Suzuki1993} was used to evolve states in time \cite{Vidal2003, White2004, Daley2004}. Alternatively, the Dirac-Frenkel time-dependent variational principle \cite{Haegeman2011, Haegeman2016} or the Krylov-based approaches \cite{GarciaRipoll2006, Dargel2011, Wall2012} can be used. Another approach is to approximate the time-evolution operator by a matrix product operator (MPO) representation, either through a first-order approximation of the Taylor series  \cite{Zaletel2015} or by means of a cluster expansion \cite{Vanhecke2021, Vanhecke2023}. This can be combined with efficient routines for applying a time-evolution MPO to a given MPS to faithfully represent a time-evolved state. Recently, an MPO encoding of the Taylor series was developed that is accurate up to arbitrary order in the time step, is size-extensive (making it applicable in the thermodynamic limit), and can handle long-range interactions \cite{VanDamme2024}.

Yet, these strategies cannot be straightforwardly applied to time-dependent Hamiltonians, so in practice, one resorts to splitting the time evolution into small time steps and keeping the Hamiltonian constant in each evolution step \cite{Kennes2018, Osterkorn2023, Gadge2025}; for rapidly fluctuating Hamiltonians, this can require the use of very small time steps and, therefore, high computational resources. Alternatively, one can approximate the Magnus expansion (or its commutator-free approximation) in terms of tensor networks and apply the usual methods for exponentiating a local operator \cite{Wall2012, Gaggioli2025}.

In the context of quantum simulation, this question takes the form of how to approximate the time evolution of a given lattice Hamiltonian in terms of a quantum circuit \cite{mckeever2023, mckeever2024}. For time-independent Hamiltonians, the Trotter-Suzuki decomposition of the time-evolution operator arises as the most straightforward option because it can often be readily implemented as a sequence of quantum gates \cite{Lloyd1996}. For time-dependent Hamiltonians, however, the situation is again more complicated \cite{Huyghebaert1990, Poulin2011}. It has been proposed to truncate the Dyson series \cite{Kieferova2019, Low2019}, as a time-dependent version of the Taylor expansion of the time-evolution operator \cite{Berry2015}. Yet, also in this context, the design of efficient quantum circuits that simulate the evolution under time-dependent Hamiltonians is an ongoing challenge.

In this work, we generalize the ideas of Ref.~\onlinecite{VanDamme2024} to the setting of time-dependent Hamiltonians. We identify two different routes to accomplish this: either we construct the Magnus operator [Eq.~\eqref{eq:magnus}] in terms of MPOs and perform the usual time-independent exponentiation, or we develop an explicit encoding of the Dyson series [Eq.~\eqref{eq:dyson-series}] as an MPO. Both approaches are shown to work for arbitrary order and for both short- and long-range interactions, and both constructions are shown to yield size-extensive operators that can be formulated on finite systems or directly in the thermodynamic limit.

The outline of this paper is as follows. We first set the MPS-MPO notation in Sec.~\ref{sec:mps_mpo}, and we recapitulate the time-independent construction from Ref.~\onlinecite{VanDamme2024} in Sec.~\ref{sec:taylor}. Then we outline the approach based on the Magnus expansion in Sec.~\ref{sec:magnus} and the one based on the Dyson series in Sec.~\ref{sec:dyson}. Then, in Sec.~\ref{sec:compression}, we show two ways to compress the Dyson MPO, which is needed to significantly reduce the bond dimension. We benchmark our Dyson MPO construction in Sec.~\ref{sec:benchmarks} and conclude with an outlook in Sec.~\ref{sec:outlook}.

\section{Matrix product states and matrix product operators}
\label{sec:mps_mpo}

Our derivations will rely heavily on the tensor network formalism \cite{Schollwoeck2011, Cirac2021, xiang2024} for describing matrix product states (MPS) and matrix product operators (MPO), as well as the representation of an MPO in terms of a finite state machine. Therefore, in this section, we will introduce these notions in a self-contained way.

\subsection{Matrix product states}
\label{sec:mps}

We consider a quantum spin chain\footnote{We can treat itinerant bosons or fermions without many modifications, as well as quasi 1-D systems such as cylinders by mapping to a 1-D geometry and introducing longer-range terms in the Hamiltonian.} of $N$ sites with local physical dimension $d$. An MPS is obtained by assigning to every site a local three-leg tensor $A_i$ (one leg corresponding to the physical spin and two virtual legs that carry the entanglement), placing them in a string-like sequence and contracting over all virtual legs. We use a graphic tensor network representation, allowing us to write the MPS as
\begin{equation}
    \ket{\Psi(\{A_i\})} = \raisebox{-0.45\height}{\includegraphics[scale=0.8,page=1]{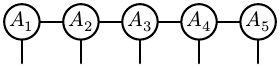}} ,
\end{equation}
where the circles denote the tensors, and connected legs indicate that these are contracted (i.e., summed over a shared index). In an MPS, the spins are locally correlated and the entanglement obeys an area law, so that the class of MPS are an excellent variational class for representing ground states of local Hamiltonians \cite{fannes1992, verstraete2006}. The dimensions of the virtual legs of the tensors are called the bond dimensions of the MPS $\{D_i\}$; the maximal value $D=\max D_i$ serves as a control parameter when approximating states by MPS. The larger the bond dimension, the more expressive the corresponding variational class of MPS.

One of the most fruitful properties of the MPS representation of states is its extensivity: by concatenating an extensive number of local tensors, we can build many-body wavefunctions for large systems. This property even allows us to represent states directly in the thermodynamic limit: we construct a state by repeating the same local unit cell of $M$ tensors $\{A_i,i=1\dots M\}$, giving rise to an extensive wavefunction for the infinite system that is translation invariant over $M$ sites. In the remainder of this section, we will restrict to the case of single-site unit cells to make the notation more succinct, in which case the MPS is parametrized by a single tensor $A$:
\begin{equation}
    \ket{\Psi(A)} = \raisebox{-0.45\height}{\includegraphics[page=2,scale=0.8]{figures/tensors.pdf}} \;.
\end{equation}
The norm of such a state is given by the infinite diagram (with $\bar{A}$ denoting the complex conjugate of the tensor $A$)
\begin{equation}
    \raisebox{-0.45\height}{\includegraphics[page=3,scale=0.8]{figures/tensors.pdf}},
\end{equation}
which can be seen as the infinite product of the operator
\begin{equation}
    T = \raisebox{-0.45\height}{\includegraphics[page=4,scale=0.8]{figures/tensors.pdf}},
\end{equation}
called the MPS transfer matrix. An MPS for which the leading eigenvalue of the transfer matrix is set to one\footnote{By construction, the transfer matrix is a completely positive map, so the leading eigenvalue is guaranteed to be positive if it is unique. The latter property is guaranteed if we restrict to injective MPS, which we will do without exception in this work (see Ref.~\onlinecite{vanderstraeten2019} for more details). Any injective MPS can therefore be normalized by simply rescaling the MPS tensors such that the leading eigenvalue of the transfer matrix is equal to one.} is, therefore, said to be normalized
\begin{equation}
    \lambda_{\max} (T) = 1 \;.
\end{equation}
Similarly, the overlap between two normalized MPS $\ket{\Psi(A)}$ and $\ket{\Psi(B)}$ in the thermodynamic limit can be characterized by the leading eigenvalue of the mixed transfer matrix
\begin{equation}
    \label{eq:mixedtransfermatrix}
    \lambda_{AB} = \lambda_{\max}(T_{AB}), \qquad T_{AB} = \raisebox{-0.45\height}{\includegraphics[page=5,scale=0.8]{figures/tensors.pdf}} \;.
\end{equation}

After having normalized the MPS, we can also compute local expectation values efficiently. A local spin measurement is represented graphically by inserting an operator on one of the sites within the unit cell
\begin{equation}
    \raisebox{-0.45\height}{\includegraphics[page=6,scale=0.8]{figures/tensors.pdf}}\;.
\end{equation}
In this diagram, we can see an infinite sequence of transfer matrix operations to the left and to the right of this local unit cell, which we can represent by the left and right fixed points of the transfer matrix:
\begin{equation}
    \raisebox{-0.45\height}{\includegraphics[page=7,scale=0.8]{figures/tensors.pdf}}, \; \raisebox{-0.45\height}{\includegraphics[page=8,scale=0.8]{figures/tensors.pdf}}
\end{equation}
corresponding to the projectors onto the leading subspace of the transfer matrix. By normalizing these fixed points as
\begin{equation}
    \raisebox{-0.45\height}{\includegraphics[page=9]{figures/tensors.pdf}} = 1,
\end{equation}
a local expectation value is finally given by
\begin{equation}
    \bra{\Psi(A)} O_i \ket{\Psi(A)} = \raisebox{-0.45\height}{\includegraphics[page=10,scale=0.8]{figures/tensors.pdf}} \;.
\end{equation}
Although we are considering states directly in the thermodynamic limit, this local  expectation value is well defined and can be evaluated with a cost scaling as $\mathcal{O}(D^3)$ in the MPS bond dimension $D$.

In particular, for a given model Hamiltonian, we can evaluate the variational energy density of the MPS. In the next step, this variational energy density can be minimized with respect to the parameters in the MPS tensors $\{A_i\}$ in order to obtain the variationally optimal state within the class of MPS with a certain bond dimension. Throughout the years, many performant algorithms have been developed \cite{Vidal2007, McCulloch2008, ZaunerStauber2018} for solving this variational optimization problem efficiently.

\subsection{Matrix product operators}
\label{sec:mpo}

In tensor network language, a matrix product operator (MPO) \cite{Verstraete2004a} is obtained by adding another physical leg to each tensor, obtaining an operator instead of a state. In diagrammatic language, we obtain for an MPO with a two-site unit cell in the thermodynamic limit
\begin{equation}
    P(\{O_i\}) = \raisebox{-0.52\height}{\includegraphics[page=11,scale=0.75]{figures/tensors.pdf}} \;.
\end{equation}
Generically, such an MPO is a size-extensive operator: by applying this MPO to an MPS, we obtain a new MPS
\begin{equation}
    \ket{\Psi(\{A_i'\})} = P(\{O_i\}) \ket{\Psi(\{A_i\})}
\end{equation}
that can be normalized. The new MPS tensors $\{A_i'\}$ are simply obtained by grouping tensor indices:
\begin{equation}
    \raisebox{-0.5\height}{\includegraphics[page=12,scale=0.8]{figures/tensors.pdf}} = \raisebox{-0.5\height}{\includegraphics[page=13,scale=0.8]{figures/tensors.pdf}} \;.
\end{equation}
The new MPS has an enlarged bond dimension and in numerical algorithms involving MPO-MPS multiplications, one typically needs to truncate the bond dimension in order to keep the sizes tractable. A range of powerful and efficient algorithms for MPS truncation has been developed over the years, including the use of singular value decompositions \cite{Schollwoeck2011}, the zip-up algorithm \cite{Stoudenmire2010, Paeckel2019}, a density-matrix based algorithm \cite{Stoudenmire}, a variation of the DMRG-algorithm \cite{Verstraete2004a, Verstraete2004b}, and variational strategies \cite{Vanhecke2021b}.

Consider now a Hamiltonian
\begin{equation} \label{eq:h}
    H = \sum_{n\in\mathcal{L}} h_n,
\end{equation}
where the sum runs over all sites in the chain $\mathcal{L}$, and where $h_n$ is either a $k$-local operator acting non-trivially on the region $(n,n+1,\dots,n+k-1)$, or a sum of exponentially decaying interactions acting on the unbounded region $(n,n+1,\dots)$. We will represent this Hamiltonian in terms of an MPO \cite{Pirvu2010},
\begin{equation}
    H = \raisebox{-0.43\height}{\includegraphics[page=14,scale=0.75]{figures/tensors.pdf}}
\end{equation}
with the MPO tensors now taking the special upper-triangular form \cite{McCulloch2007, Crosswhite2008}
\begin{equation}
     \label{eq:jordan_mpo}
    \raisebox{-0.45\height}{\includegraphics[page=18, scale=0.8]{figures/tensors.pdf}}  = \raisebox{-0.375\height}{\includegraphics[page=2]{figures/matrices.pdf}} \;.
\end{equation}
Here we view the MPO tensor as a matrix labelled by the ``levels'' of its virtual legs (in red and blue), where we have denoted the virtual levels of the MPO tensors explicitly as $(1,2,3)$. The levels $(1)$ and $(3)$ have dimension one, but the middle level $(2)$ can generally take a larger value $\chi$ \footnote{We refer to \(\chi\) as the bond dimension of the Hamiltonian MPO. In reality the total bond dimension of the MPO is \(2 + \chi\).}. Therefore, $L$ and $R$ are respectively an operator-valued row and column vector of length $\chi$, whereas $A$ is an operator-valued matrix of dimensions $(\chi\times\chi)$ and $D$ is a single-site operator. These MPO tensors are combined with (scalar valued) boundary vectors of the form
\begin{equation}
v_L = \begin{pmatrix}1 & 0 & 0 \end{pmatrix},\quad v_R = \begin{pmatrix} 0 \\ 0 \\ 1 \end{pmatrix}
\end{equation}
such that contracting the virtual levels gives rise to a Hamiltonian of the form in Eq.~\eqref{eq:h} with
\begin{multline}
    h_n = D_n + L_n\cdot R_{n+1} + L_n \cdot A_{n+1} \cdot R_{n+2} \\ + L_n \cdot A_{n+1} \cdot A_{n+2} \cdot R_{n+3} + \dots \;.
\end{multline}
This MPO can also be understood by use of the finite-state machine (FSM) representation of MPOs \cite{Crosswhite2008, Parker2020}. The Hamiltonian can for example be represented as
\begin{equation} \label{eq:fsm_h}
    \begin{aligned}
        \includegraphics[page=7,scale=0.8]{figures/fletcher-graphs.pdf}
    \end{aligned}
\end{equation}
When applying the Hamiltonian on a string of sites from left to right, we move through this diagram from left to right, applying the operator written along the edge of the transition path.

Crucially, such an MPO with an upper-triangular form as in Eq.~\eqref{eq:jordan_mpo} is \emph{not} a size-extensive MPO in the following sense: when applying this MPO to an MPS, the new MPS is no longer normalizable and it can therefore not be truncated by the usual methods. Instead, its expectation value with respect to an MPS scales with system size $N$ as
\begin{equation}
    \bra{\Psi(A)} H \ket{\Psi(A)} \propto N
\end{equation}
as it should for a local Hamiltonian. Because of this scaling, this MPO is said to be a first-degree MPO \cite{Parker2020}, to make explicit the distinction between a generic (zeroth-degree) MPO that can be applied to an MPS. The MPO construction for representing a (quasi) local Hamiltonian should therefore be seen as a convenient bookkeeping device for summing up Hamiltonian terms in tensor network algorithms, rather than an extensive operator that can be applied to an MPS.

\subsection{Operations on first-degree MPOs}
\label{sec:operations}

We can efficiently perform operations on these first-degree MPOs such as additions and multiplications \cite{Parker2020}. Indeed, two Hamiltonians $H_1$ and $H_2$, both represented by an MPO of the form in Eq.~\eqref{eq:jordan_mpo} with entries $\{L^{(i)},R^{(i)},A^{(i)},D^{(i)}\}_{i=1,2}$, can be added to give a Hamiltonian $H=H_1+H_2$ represented by an MPO with tensors
\begin{equation} \label{eq:sum}
    H \sim \begin{pmatrix} \one & L^{(1)} & L^{(2)} & D^{(1)}+D^{(2)} \\
    & A^{(1)} & 0 & R^{(1)} \\
    & 0 & A^{(2)} & R^{(2)} \\
    & & & \one
    \end{pmatrix} \;,
\end{equation}
which is again in the same upper-triangular form, i.e.~a first-degree MPO. Multiplying a Hamiltonian $H$ represented by $\{L,R,A,D\}$ by a scalar $\lambda$ similarly gives rise to a Hamiltonian $H'=\lambda H$ represented by
\begin{equation}
    H' \sim \begin{pmatrix} \one & \lambda L  & \lambda D \\
    & A & R \\
    & & \one
    \end{pmatrix}.
\end{equation}
Note that we could have chosen to multiply $L$ and $R$ with any two factors of a factorisation of $\lambda$ instead (e.g. $\sqrt\lambda$ and $\sqrt\lambda$) .

The multiplication of two Hamiltonians $H^{(1)}$ and $H^{(2)}$, each represented by a first degree MPO, requires a bit more work. Straightforwardly concatenating the two MPO tensors
\begin{equation}
    H^{(1)} H^{(2)} = \raisebox{-0.45\height}{\includegraphics[page=16]{figures/tensors.pdf}}
\end{equation}
gives rise to a two-index structure of the virtual level. In terms of an FSM, this MPO is represented as the diagram
\begin{equation}
    \label{eq:H2diagram}
   \raisebox{-0.46\height}{\includegraphics[page=5,scale=0.75]{figures/fletcher-graphs.pdf}} \;.
\end{equation}
However, this no longer represents a first-degree MPO. Indeed, if \(H\) is a sum of \(\mathcal{O}(N)\) local terms, \(\langle H^n\rangle\) contains \(\mathcal{O}(N^n)\) expectation values of local terms \cite{Parker2020}. Similarly, starting from a normalized state $\ket{\Psi_0}$, acting with powers of the Hamiltonian gives rise to a state that is not normalizable,
\begin{equation} \label{eq:extensivity}
    \ket{\Psi_n} = H^n \ket{\Psi_0} \quad\Rightarrow\quad \braket{\Psi_{n'}|\Psi_n} \propto N^{n+n'} \;.
\end{equation}
The reason for this higher-order scaling is situated in the $(1,3)$ and $(3,1)$ levels in the FSM representation in Eq.~\eqref{eq:H2diagram}. When running through the FSM, the level $(1,3)$ represents the situation where a $H^{(1)}$ term has already been applied and a $H^{(2)}$ term is yet to start: it connects the non-overlapping or disconnected actions of terms in $H^{(1)}$ and $H^{(2)}$, and clearly the number of such terms scales quadratically with system size.
To capture this, we propose to write the product of two Hamiltonians as the sum of two contributions
\begin{equation}
    H^{(1)} H^{(2)} = (H^{(1)} H^{(2)})_{\odot} + (H^{(1)}H^{(2)})_{\otimes} \;.
\end{equation}
The first term is the non-disjoint product \((\cdot)_{\odot}\), containing all terms that overlap on at least one site \cite{Parker2020}, whereas the second term is the disjoint product \((\cdot)_{\otimes}\), containing all non-overlapping terms. The MPO of the non-disjoint product \((H^{(1)}H^{(2)})_{\odot}\) is represented by the FSM where the $(1,3)$ and $(3,1)$ levels have been omitted:
\begin{equation}
    \label{eq:H12disjoint}
   \raisebox{-0.385\height}{\includegraphics[page=6,scale=0.75]{figures/fletcher-graphs.pdf}} \;.
\end{equation}
With the problematic levels removed, this now represents an MPO of the form
\begin{equation} \label{eq:H1H2_non-disjoint}
(H^{(1)}H^{(2)})_\odot \sim \begin{pmatrix} \one & L^{(12)}  & D^{(12)} \\
    & A^{(12)} & R^{(12)} \\
    & & \one
    \end{pmatrix}
\end{equation}
with the entries
\begin{align}
    L^{(12)} &= \begin{pmatrix} L^{(1)} & L^{(2)} & L^{(1)}L^{(2)} & L^{(1)}D^{(2)} & D^{(1)} L^{(2)} \end{pmatrix} \nonumber \\
    R^{(12)} &= \begin{pmatrix} R^{(1)}D^{(2)} \\ D^{(1)} R^{(2)} \\ R^{(1)} R^{(2)} \\ R^{(1)} \\ R^{(2)}  \end{pmatrix} \nonumber  \\
    A^{(12)} &= \begin{pmatrix}
    A^{(1)} & & A^{(1)}L^{(2)} & A^{(1)}D^{(2)}& R^{(1)}L^{(2)} \\
    & A^{(2)} & L^{(1)}A^{(2)} & L^{(1)}R^{(2)}& D^{(1)}A^{(2)} \\
    & & A^{(1)}A^{(2)} & A^{(1)}R^{(2)} & R^{(1)}A^{(2)} \\
    & & & A^{(1)} & \\
    & & & & A^{(2)}
    \end{pmatrix} \nonumber \\
    D^{(12)} &= D^{(1)}D^{(2)} \;.  \label{eq:H1H2_entries}
\end{align}
Here, juxtaposition of two MPO entries (e.g. \(D^{(1)}D^{(2)}\)) corresponds to a product on the physical level, and a tensor product on the virtual level. In this form, it is manifest that the non-disjoint product of two first-degree MPOs is itself a first degree MPO with a bond dimension
\begin{equation}
    \chi_{12} = 2\chi_1 + 2\chi_2 + \chi_1\chi_2 \;.
\end{equation}

This disjoint product is now particularly useful for computing the commutator between two Hamiltonians. Indeed, we observe
\begin{align}
    [H^{(1)},H^{(2)}] = (H^{(1)}H^{(2)})_{\odot} - (H^{(2)}H^{(1)})_{\odot}\;,
\end{align}
because the non-overlapping contributions in $H^{(1)}H^{(2)}$ and $H^{(2)}H^{(1)}$ are identical \cite{Parker2020}. With the right-hand side being a linear combination of two first-degree MPOs, the commutator between two Hamiltonians can be computed straightforwardly, itself in the form of a first-degree MPO.

For the special case of the non-disjoint square of a first-degree MPO $H^{(1)}$, the representation can be compressed. In $L^{(11)}$ the entry $L^{(1)}$ appears twice, so we can simply rewire all connections to one of the two instances of $L^{(1)}$ and omit the column with the other instance. The rewiring amounts to summing two rows in $A^{(11)}$ and $R^{(11)}$. Similarly, we can omit one of the two instances of the entry $R^{(1)}$ in $R^{(11)}$, and add the corresponding columns in $A^{(11)}$ and $L^{(11)}$. These are special cases of a more generally applicable compression strategy discussed in Section~\ref{sec:compression}. The result of this compression is the following block structure for the non-disjoint square $(H^{(1)}H^{(1)})_\odot$:
\begin{align}
    L^{(11)} &= \begin{pmatrix} L^{(1)} & L^{(1)}L^{(1)} & \{D^{(1)},L^{(1)}\} \end{pmatrix} \nonumber \\
    R^{(11)} &= \begin{pmatrix} \{D^{(1)},R^{(1)}\} \\ R^{(1)}R^{(1)} \\ R^{(1)} \end{pmatrix} \nonumber  \\
    A^{(11)} &= \begin{pmatrix} A^{(1)} & \{A^{(1)},L^{(1)}\} & \{L^{(1)},R^{(1)}\}+\{A^{(1)},D^{(1)}\} \\ & A^{(1)}A^{(1)} & \{A^{(1)},R^{(1)}\} \\ & & A^{(1)} \end{pmatrix} \nonumber \\
    D^{(11)} &= D^{(1)}D^{(1)} \;.\label{eq:H1H1_compressed}
\end{align}

Here we have introduced the notation of an anticommutator of two MPO entries, for example:
\begin{equation}
    \{A,L\} = AL + LA = \raisebox{-0.45\height}{\includegraphics[page=17, scale=0.75]{figures/tensors.pdf}}
\end{equation}
and similarly for the other combinations. The MPO bond dimension for the non-disjoint square can therefore be reduced to $\chi_{11}=2\chi_1+\chi_1^2$.

Finally, we note that a canonical form exists for these first-degree MPOs, which can then be used to truncate the MPO bond dimension accurately \cite{Parker2020}. All of the above constructions and manipulations can therefore be combined with truncation steps.

\section{The Taylor series as an MPO}
\label{sec:taylor}

This section reviews the MPO construction for taking the exponential of a generic local Hamiltonian, represented by a first-degree MPO, from Ref.~\onlinecite{VanDamme2024}, in order to set the stage for our MPO construction of the Dyson series in Sec.~\ref{sec:dyson}. Hereto, we start from the Taylor expansion of the evolution operator \begin{samepage}$U(t,t_0) = \exp(-i (t-t_0) H)$\end{samepage} in the case of a time-independent Hamiltonian, which we write as
\begin{equation}
     \e^{\tau H} =  \one + \tau H + \frac{\tau^2}{2!} H^2 + \frac{\tau^3}{3!} H^3 +  \dots \;.
\end{equation}
with $\tau = -i(t-t_0)$. Because the powers of $H$ have non-matching scaling as a function of system size [Eq.~\eqref{eq:extensivity}], applying them to a state will yield a state that cannot be properly and intrinsically normalized in the thermodynamic limit. Hence, the construction of Ref.~\onlinecite{VanDamme2024} reorganizes the different contributions in the Taylor expansion to obtain a size-extensive representation at each order. This procedure took inspiration from the size-extensive representations of perturbative series expansions \cite{Vanderstraeten2017, Chen2022} or cluster expansions \cite{Vanhecke2021, Vanhecke2023} in terms of tensor network operators, and is a generalization of the so-called $W_{\text{I}}$ operator that was introduced in Ref.~\onlinecite{Zaletel2015}.

\subsection{First-order Taylor MPO}

The crucial ingredient of this construction is a transformation that turns a first-degree MPO back into an extensive MPO. This transformation can be visualized in terms of a FSM diagram as
\begin{equation}
    \raisebox{-0.375\height}{\includegraphics[page=7, scale=0.75]{figures/fletcher-graphs.pdf}} \rightarrow \raisebox{-0.375\height}{\includegraphics[page=8, scale=0.75]{figures/fletcher-graphs.pdf}} \;.
\end{equation}
Hence, every arrow arriving at level $(3)$ is rerouted back to level $(1)$, and the corresponding entry is multiplied by a factor $\tau = -i(t-t_0)$. Put differently: after applying a term in the Hamiltonian, instead of arriving on the level $(3)$, we go back to the level $(1)$ with the given factor. In practice this amounts to multiplying the column labelled by $(3)$ with $\tau$ and adding it to the $(1)$ column. The $(3)$ column and row are then deleted. This transformation therefore gives rise to a new MPO with tensor entries
\begin{equation}
    O(\tau) = \begin{pmatrix} \one + \tau D & L \\ \tau R & A \end{pmatrix} \;.
\end{equation}
The validity of this transformation can be confirmed by taking the derivative of the resulting MPO at $\tau=0$, upon which the first-degree MPO representation of $H$ is recovered, as explained in Appendix~\ref{app:MPSderivatives}. Thus, $O(\tau)$ encodes an MPO $W(O)$ that satisfies $\left.W(O)\right|_{\tau=0} = \one$ and $\left.\frac{d\ }{d \tau} W(O)\right|_{\tau=0} = H$. Furthermore, in the absence of the upper-diagonal structure, this MPO is now a size-extensive operator:  Applying the MPO to a normalized state gives rise to a state for which the norm scales exponentially with system size. The extensivity of the MPO has the added benefit that it encodes all higher-order terms of the Taylor expansion that can be written as the product of non-overlapping first-order terms. Indeed, we find that $O(\tau)$ encodes the operator
\begin{equation}
    W(O) = \one + \tau H + \frac{\tau^2}{2!} (HH)_{\otimes} + \frac{\tau^3}{3!} (HHH)_{\otimes} + \dots,
\end{equation}
in terms of the disjoint powers of the Hamiltonian.

\subsection{Higher orders}

\begin{algorithm}[t]
\caption{$N$-th order Taylor MPO}
\label{alg:Taylor}
\textbf{Input:} $H$, $\tau$, $N$ \\
\textbf{Output:} $O$ (Taylor MPO of $N$-th order)\\
\(O\leftarrow H^N\)\\
\For{\(l \in\) levels of  \(O\)}{
    \(n_2 \leftarrow \) Amount of 2's in $l$\\
    \(n_3 \leftarrow \) Amount of 3's in $l$\\
    \If{\((n_2 == 0)\) \(\wedge\) \((1\le n_3)\)}{
    $O[:, 1] = O[:, 1]+ \tau^{n_3} \cdot \frac{ (N-n_3)!}{N!}O[:, l]$\\
    Remove level $l$
    }
}
\Return O
\end{algorithm}

This expression immediately tells us which terms are missing from the MPO in order to approximate the Taylor expansion to higher orders. Indeed, these are precisely the terms in $H^2$ in which two Hamiltonian operators overlap, contained in the non-disjoint product $(HH)_{\odot}$. We can do a similar transformation that turns \(H^2\) into an extensive MPO. The problematic levels that keep \(H^2 \) from being an extensive operator are \((13), (31)\) and \((33)\). We again reroute these levels to \((11)\), now multiplying with the correct power of \(\tau\). We multiply columns \((13)\) and \((31)\) by $\tau/2$ (the factor 1/2 arises in order to avoid double counting), and column \((33)\) by $\tau^2/2$, and add them to the first column. Then we remove those rows and columns from the MPO. We obtain a second-order Taylor MPO\footnote{Note that second-order in this context means that the MPO captures all terms of the Taylor series up to \(\tau^2\). It is not a second-degree MPO, because its norm scales exponentially with the system size.} of the form
\begin{equation} \label{eq:taylor_order2}
\begin{pmatrix}
        \one + \tau D^{(1)} + \frac{\tau^2}{2} D^{(11)} & L^{(1)} & L^{(11)} \\
        \tau R^{(1)} & A^{(1)} & 0\\
         \frac{\tau^2}{2} R^{(11)} & 0 & A^{(11)}
    \end{pmatrix}\;.
\end{equation}
where the blocks $L^{(11)}$, $R^{(11)}$, $D^{(11)}$ and $A^{(11)}$ already appeared in the non-disjoint square of the first-degree MPO $H^{(1)}$ [Eq.~\eqref{eq:H1H1_compressed}].

These blocks capture both the first and second order contributions to the Taylor expansion. Note again the extensivity property of the resulting MPO: now it not only captures the fully disconnected third order terms (which we already had in the first-order MPO), but also all third-order terms that consist of a second-order overlapping part and a disconnected first-order part, with the correct prefactor:
\begin{multline}
    \one + \tau H^{(1)} + \frac{\tau^2}{2} (H^{(1)})^2 + \frac{\tau^3}{3!} \left(H^{(1)}H^{(1)}H^{(1)}\right)_\otimes \\ + \frac{\tau^3}{3!} \left( (H^{(1)}H^{(1)})_\odot H^{(1)}\right)_\otimes + \mathcal{O}(\tau^4) \;.
\end{multline}
The only missing third-order contribution is
\begin{equation}
    ((H^{(1)}H^{(1)})_{\odot}H^{(1)})_\odot,
\end{equation}
i.e.\ terms where all three Hamiltonian operators overlap.

This procedure can now be algorithmically generalized to higher orders. A key observation is that, by constructing the MPO representation of $H^N$, it also exactly encodes the all-3 columns of lower powers $H^k$ for $k=1,2,\ldots$, namely in these levels $l$ with $k=n_3$ indices $3$ in their multi-index structure, and $n_1 = N-n_3$ indices $1$. These levels, of which there are $\frac{N!}{n_3! (N-n_3)!}$ and which are all equivalent, should thus not be simply discarded as in the case of constructing $H^{\odot N}$, but the corresponding columns should be added to the level $l=1=(1,1,\ldots,1)$ with the associated factor $\frac{\tau^k}{\cancel{n_3!}} \frac{\cancel{n_3!} (N-k)!}{N!}$.
This leads to the algorithm outlined in Alg.~\ref{alg:Taylor}, which is implemented in the \texttt{MPSKit.jl} software package \cite{MPSKit}.

The resulting MPO approximation can be improved in several ways. Firstly, the bond dimension can be reduced by implementing exact and approximate compressions steps, as was already discussed in  Ref.~\onlinecite{VanDamme2024}, which also lists explicit forms for the compressed and extended first- and second-order Taylor MPOs. These original compression strategies can further be enhanced and generalized, so that they become applicable to the MPO approximation of the Dyson series, as discussed below in Section~\ref{sec:compression} and Appendix~\ref{app:rowcompression}. When these strategies are applied to the Taylor MPO, they yield a further reduction of bond dimension compared to Ref.~\onlinecite{VanDamme2024}, giving rise to the results summarized in Table~\ref{tab:compression} in Appendix~\ref{app:rowcompression}. Furthermore, as we will show in a forthcoming paper \cite{dtermupcoming}, the contribution of on-site (``$D$'') terms can be included separately and up to significantly higher order.

\section{The Magnus expansion as an MPO}
\label{sec:magnus}

Let us now consider time-dependent Hamiltonians. We will consider translation-invariant Hamiltonians $H(t)$, which we decompose in their different ``driving channels'' $a = \{1, 2, \dots\}$, as the sum of time-independent local or quasi-local hamiltonians $H^{(a)}$ multiplied by time-dependent coefficients, i.e.\ ``driving functions'' $f_a(t)$:
\begin{equation}\label{eq:TD-Hamiltonian}
    H(t) = \sum_a f_a(t) H^{(a)}
 \;.\end{equation}
In this decomposition, every $H^{(a)}$ is an extensive Hamiltonian of the form of Eq.~\eqref{eq:h}
\begin{equation}
H^{(a)} = \sum_{n\in\mathcal{L}} h^{(a)}_{n}
\end{equation}
which can be represented as first-degree MPOs with entries $\{L^{(a)}, R^{(a)}, A^{(a)}, D^{(a)}\}$. All time dependence is now contained in the driving functions $f_a(t)$.

In this section, the strategy for approximating the time-evolution operator $U(t,t_0)$ up to given order $N$ is to (i) write down the Magnus operator $\Omega(t,t_0)$ [Eq.~\eqref{eq:magnus_operators}] up to order $N$ as a first-degree MPO, and (ii) apply the extensive Taylor procedure up to the same order to exponentiate it. Note that a brute-force addition of all the different terms in the Magnus expansion up to a given order $N$ can be done for finite systems, but quickly leads to very high bond dimensions \cite{Gaggioli2025}. Moreover, for an infinite system we have shown in Sec.~\ref{sec:mps_mpo} that adding different powers of the Hamiltonian is ill-defined in the thermodynamic limit. A crucial insight, however, is that the commutators appearing in the Magnus expansion can actually all be written as first-degree MPOs, as we have already noted in Sec.~\ref{sec:operations}.

Let us start with the first term in the Magnus expansion, which is just the integral of the Hamiltonian over a time step. Simply integrating the Hamiltonian yields
\begin{align}
    \Omega_1(t,t_0) &= -i \int_{t_0}^t H(t_1) dt_1 \nonumber \\
    &= \sum_a [f_a] H^{(a)}
\end{align}
where we have introduced the notation for integrating a driving function over the time interval:
\begin{equation} \label{eq:bracket_f}
    [f] = - i \int_{t_0}^t f(t_1) dt_1  \;.
\end{equation}
The first Magnus operator is therefore a linear combination of first-degree MPOs with prefactors given by the integrals of the driving functions $f_a(t)$. This linear combination can be done efficiently using the tools from Sec.~\ref{sec:operations}. For the case of two driving channels in the hamiltonian $H^{(1)}$ and $H^{(2)}$ with driving functions $f_1(t)$ and $f_2(t)$, we can simply use the expression for the sum of two first-degree MPOs in Eq.~\eqref{eq:sum}
\begin{equation}
    \Omega_1(t,t_0) \sim \begin{pmatrix} \one & L^{(1)} & L^{(2)} & [f_1] D^{(1)} + [f_2] D^{(2)} \\
    & A^{(1)} & 0 & [f_1] R^{(1)} \\
    & 0 & A^{(2)} & [f_2] R^{(2)} \\
    & & & \one
    \end{pmatrix}
\end{equation}

The second order term in the Magnus expansion is:
\begin{align}
    \Omega_2(t,t_0)= -\frac{1}{2}\int_{t_0}^t dt_1\int_{t_0}^{t_1} dt_2 [H(t_1), H(t_2)],
\end{align}
which we write out as
\begin{equation}
    \Omega_2(t,t_0) = \frac{1}{2} \sum_{ab}  [f_af_b] [H^{(a)},H^{(b)}]
\end{equation}
with the notation for the integrated driving functions
\begin{equation} \label{eq:bracket_fg}
    [fg] = (-i)^2\int_{t_0}^t dt_1 \int_{t_0}^{t_1} dt_2 f(t_1)g(t_2) \;.
\end{equation}
These time-ordered integrals can be calculated efficiently with their quantic tensor train representation, as detailed in Appendix \ref{sec:timeorderedintegrals}. Based on the expression for the non-disjoint commutator of two first-degree MPOs in Eq.~\eqref{eq:H1H2_non-disjoint}, we write down an explicit MPO representation for $\Omega_2(t,t_0)$:
\begin{equation}
    \label{eq:second-order-magnus}
    \begin{pmatrix} \one & L^{(12)} & L^{(21)} & \alpha\cdot D^{(12)} - \alpha\cdot D^{(21)} \\
    & A^{(12)} & 0 & \alpha\cdot R^{(12)} \\
    & 0 & A^{(21)} & -\alpha\cdot R^{(21)} \\
    & & & \one
    \end{pmatrix} \;,
\end{equation}
where \(\alpha=\frac{1}{2}([f_1f_2]-[f_2f_1])\). Eq.~\eqref{eq:second-order-magnus} is again a first-degree MPO.

This procedure can be extended to higher orders: in all orders, the Magnus operators are composed of commutators of Hamiltonian terms, and can therefore be written as linear combinations of first-degree MPOs. The bond dimension increases linearly with the number of commutators, which itself increases exponentially in the expansion order. For specific examples, it is very likely that the resulting first-degree MPOs can be rewritten with a smaller bond dimension; the truncation procedure for first-degree MPOs of Ref.~\onlinecite{Parker2020} is therefore expected to be very useful as an intermediary step in this construction.

In a next step, the Magnus operator $\Omega(t,t_0)$ can be exponentiated using the Taylor expansion MPO construction from Sec.~\ref{sec:taylor} with time step $\tau=1$. For the example of two terms, taking a first-degree Taylor expansion of the first-order Magnus operator $\Omega_1(t,t_0)$ we find
\begin{equation} \label{eq:magnus_taylor_first}
    \begin{pmatrix} \one + [f_1] D^{(1)} + [f_2] D^{(2)} & L^{(1)} & L^{(2)} \\
    [f_1] R^{(1)} & A^{(1)} & 0  \\
    [f_2] R^{(2)} & 0 & A^{(2)}  \\
    \end{pmatrix} \;.
\end{equation}
In general, one can write down a Taylor expansion of order $N$ of a Magnus operator of order $N'$, where $N$ and $N'$ do not have to coincide in principle. However, in order to obtain all terms in the time-evolution operator up to a given order in $(dt)^M$, $N$ and $N'$ have to be both at least $M$. We will not provide any expressions for these higher-order cases, because the size of the MPOs increases dramatically.

The two-step nature of this approach -- first expanding the Magnus operator, then exponentiating it via a Taylor series -- introduces a redundancy: combining two independent expansion schemes generates higher-order terms beyond what is strictly required to achieve a given accuracy in \(\tau\), inflating the bond dimension unnecessarily. This motivates a more direct approach. While the Magnus expansion as a first-degree MPO remains a powerful tool in its own right, our focus here is on time-evolution. We therefore turn to a direct expansion of the time-ordered evolution operator via the Dyson series, which avoids this overhead entirely.

\section{The Dyson series as an MPO}
\label{sec:dyson}

We return to the form in Eq.~\eqref{eq:TD-Hamiltonian} of a time-dependent Hamiltonian as a sum of different non-commuting terms with scalar driving functions $f_i(t)$. From this form of $H(t)$, the Dyson series can be rephrased as:
\begin{multline}\label{eq:Dyson-Series-Written-Out}
     U(t, t_0) = \mathds{1} + \sum_a[f_a]H^{(a)} \\ + \sum_{ab}[f_a f_b] H^{(a)} H^{(b)} + \dots,
\end{multline}
with the prefactors given by the time-ordered integrals of the driving functions in Eqs. \eqref{eq:bracket_f} and \eqref{eq:bracket_fg}.

Note that this expansion generalizes the extensivity property of the time-independent case in the sense that all disjoint contributions at higher orders can be written as the disjoint products of the lower-order terms. For the disjoint contributions in second order, for example, we find
\begin{align}
    \sum_{ab} [f_af_b] & (H^{(a)}H^{(b)})_\otimes \nonumber \\
    &= \sum_a [f_af_a] (H^{(a)}H^{(a)})_\otimes\\
    & \qquad + \sum_{a<b} ([f_af_b]+[f_bf_a]) (H^{(a)}H^{(b)})_\otimes \nonumber \\
    & = \frac{1}{2} \sum_{ab} [f_a][f_b] (H^{(a)}H^{(b)})_\otimes \;,\label{eq:dyson_extensive}
\end{align}
because of the factoring property of the time-ordered integrals
\begin{equation}
    [f_af_b]+[f_bf_a] = [f_a][f_b] \;,
\end{equation}
and the fact that \((H^{(a)}H^{(b)})_{\otimes} = (H^{(b)}H^{(a)})_{\otimes}\). This extensivity property of the Dyson expansion motivates us to generalize the extensive MPO transformation that we proposed in Sec.~\ref{sec:taylor} for representing the time evolution operator of a time-independent Hamiltonian.

Note, however, that the Dyson series contains not just different powers of the time step as in the time-independent case, but now contains time-ordered integrals over all different permutations of the driving functions in the Hamiltonian. This means that we need to be able to distinguish between the different permutations of the driving functions. Where previously the second order terms would need to be multiplied by $\frac{\tau^2}{2!}$, they now need to be multiplied by the correct time-ordered integral over the correct driving functions corresponding to the specific operator strings.

\subsection{First order Dyson MPO}

In order to account for this complication, we first propose a rewiring of the MPO representation of the time-dependent Hamiltonian, where we introduce a separate level $(3_a)$ for every driving channel $a$ with the driving prefactors $f_a(t)$. For a time-dependent Hamiltonian consisting of just two terms $H^{(1)}$ and $H^{(2)}$ with driving functions $f_1(t)$ and $f_2(t)$, the rewired Hamiltonian is written as
\begin{equation} \label{eq:rewire}
    \tilde{H}(t) \sim \begin{pmatrix} \one & L^{(1)} & L^{(2)} & f_1(t) D^{(1)} & f_2(t) D^{(2)} \\ & A^{(1)} & 0 & f_1(t) R^{(1)} & \\ & 0 & A^{(2)} & &  f_2(t)R^{(2)} \\ & & & \one & \\ & & & & \one
    \end{pmatrix}
\end{equation}
In terms of the finite state machine representation, this rewiring looks like:
\begin{multline}
 \raisebox{-0.5\height}{\includegraphics[page=1, width=.55\linewidth]{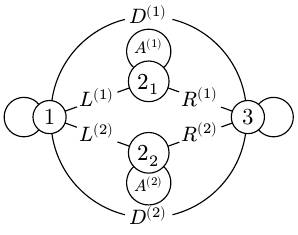}} \\ \rightarrow\qquad  \raisebox{-0.39\height}{\includegraphics[page=2, width=.55\linewidth]{figures/fletcher-graphs.pdf}} \;.
\end{multline}

In Sec.~\ref{sec:taylor} we proposed a transformation to arrive at an extensive MPO that captures the Taylor series of the evolution operator of a time-independent Hamiltonian. Here, with the rewiring into the different $(3_a)$ levels, we propose a similar transformation to encode an MPO for the first-order contributions of the Dyson series: we redirect every arrow arriving at the $(3_a)$ level back to level $(1)$ and integrate the corresponding driving function $f_a(t)$ over the time region so as to obtain $[f_a]$. For the same case of just two terms, the resulting MPO follows from Eq.~\eqref{eq:rewire}:
\begin{equation} \label{eq:dyson_first}
    \begin{pmatrix} \one + [f_1]D^{(1)} + [f_2]D^{(2)} & L^{(1)} & L^{(2)} \\
    [f_1]R^{(1)} & A^{(1)} & \\ [f_2] R^{(2)} & & A^{(2)}
    \end{pmatrix} \;.
\end{equation}
Note that this MPO is entirely equivalent to the first order Magnus operator exponentiated using the first-order Taylor expansion from Eq.~\eqref{eq:magnus_taylor_first}.

The absence of an upper-triangular structure makes this MPO size-extensive and can therefore be applied to an MPS directly in the thermodynamic limit. Moreover, all higher order terms that are made up of disconnected first-order terms are also encoded into it with the correct prefactors. Indeed, the MPO in Eq.~\eqref{eq:dyson_first} is an MPO representation of:
\begin{equation}
    \mathds{1} + \sum_a [f_a]H^{(a)} + \frac{1}{2} \sum_{ab} [f_a][f_b](H^{(a)}H^{(b)})_{\otimes} + \dots \;,
\end{equation}
which indeed captures the correct extensivity property of the Dyson expansion as in Eq.~\eqref{eq:dyson_extensive}.

\subsection{Higher orders}
\label{sec:dyson-second-order}

From the last equation, it is apparent what is missing from the MPO in order to represent the Dyson series up to second order, namely the contributions
\begin{equation}
    \sum_{ab} [f_af_b] (H^{(a)} H^{(b)})_\odot \;.
\end{equation}
An explicit expression for each of these terms can be found from Eq.~\eqref{eq:H1H2_non-disjoint}, where the non-disjoint product of two first-degree operators was written out. Similarly as we did for the second-order term in the Taylor expansion, we now start from the MPO representation of \(H^2\). We identify all columns that are labelled by levels containing only $1$s and $3_a$s, multiply them by the correct factor and add them to the first column. Afterwards, we remove those rows and columns. The correct prefactor is now prescribed by the subscript(s) of the 3-level(s) in the column label. Columns \((13_a)\) and \((3_a1)\) are to be multiplied by the factor $[f_a]/2$ (again, the factor of 1/2 is introduced to avoid double counting). The columns labelled by \((3_a3_b)\) are multiplied by the factor \([f_af_b]\).

For the case of two driving channels, there are four terms $(H^{(a)}H^{(b)})_\odot$ ($a,b=1,2$) that need to be added to the MPO. Recycling the notation from Eq.~\eqref{eq:H1H2_non-disjoint}, the full expression of the second-order Dyson MPO is given by:
\begin{equation}
    \begin{pmatrix}
        \one + \mathcal{D} & L^{(1)} & L^{(2)} & L^{(11)} & L^{(22)} & L^{(12)} & L^{(21)} \\
        [f_1] R^{(1)} & A^{(1)} & & & & & \\
        [f_2] R^{(2)} & & A^{(2)} & & & & \\
        [f_1 f_1] R^{(11)} & & & A^{(11)} & & & \\
        [f_2 f_2] R^{(22)} & & & & A^{(22)} & &\\
        [f_1f_2]R^{(12)} & & & & & A^{(12)} & \\
        [f_2f_1]R^{(21)} & & & & & & A^{(21)} \\
    \end{pmatrix}\;,
    \end{equation}
where we have reused the entries given in Eqs.~\eqref{eq:H1H2_entries} and \eqref{eq:H1H1_compressed}, and we have defined
\begin{equation}
\mathcal{D} = \sum_a [f_a]D^{(a)} + \sum_{ab} [f_af_b] D^{(a)} D^{(b)} \;.
\end{equation}
This expression for the second-order Dyson MPO is the direct generalization of the second-order Taylor MPO that was given in Eq.~\eqref{eq:taylor_order2}: the block structure is similar, but because the different driving channels give rise to non-commuting combinations and have different prefactors $[f_af_b]$, four different blocks are needed.\footnote{For a single driving channel $H^{(1)}$ with driving function $f_1(t)$, the generalization of the Taylor expansion is entirely straightforward: the prefactors $\{\tau$,$\tau^2/2,\dots\}$ are just replaced by the proper integrations of the driving function $\{[f_1],[f_1f_1],\dots\}$.}

This MPO construction can be generalized to any order $N$, and is summarized in Algorithm \ref{alg:Dyson}. Calculating the time-ordered integrals for high-order Dyson MPOs can become quite computationally expensive with traditional quadrature methods. However, they can be computed efficiently with their quantics representation \cite{waintal2026}, the details of which are explained in Appendix~\ref{sec:timeorderedintegrals}.

\begin{algorithm}[t]
\caption{$N$-th order Dyson MPO}
\label{alg:Dyson}
\textbf{Input:} $\Tilde{H}$, $(f_1, \dots, f_n)$, $(t, t_0)$, $N$ \\
\textbf{Output:} $O$ (Dyson MPO of $N$-th order)\\
\(O\leftarrow \Tilde{H}^N\)\\
\For{\(l \in\) levels of  \(O\)}{
    \(n_2 \leftarrow \) Amount of 2's in $l$\\
    \(n_3 \leftarrow \) Amount of 3's in $l$\\
    \If{\((n_2 == 0)\) \(\wedge\) \((n_3 \ge 1)\)}{
        \(\sigma \leftarrow \) Subscripts of the 3-levels in $l$\\
        \(I \leftarrow [f_{\sigma(1)} f_{\sigma(2)} \dots f_{\sigma(n_3)}]\)\\
        \(O[:, 1] = O[:, 1] + I \cdot \frac{n_3!(N-n_3)!}{N!}O[:, l]\)\\
        Remove level \(l\)
    }
}
\Return \(O\)
\end{algorithm}

\subsection{Illustration}
\label{sec:Dyson-Illustration}

As an illustration, let us focus on a simple Hamiltonian of the form
\begin{equation}
    H(t) = f(t)H^{(1)} + g(t)H^{(2)}
\end{equation}
where \(H^{(1)}\), is a sum of two-body operators (e.g. \(\sum_i L_i\otimes R_{i+1}\)) and \(H^{(2)}\) is a sum of one-site operators (e.g. \(\sum_i D_i\)).

The term $LD\otimes R$ in the Dyson series, which is an overlapping second-order term, should be multiplied with $[fg]$.
Because of the new $3$-levels that were introduced into the Hamiltonian, we can identify which time-ordered integral each term has to be multiplied by. The $R$ of $LD\otimes R$ is in column $(3_13_2)$ of \(\Tilde{H}^2\), which tells us that this term needs to be multiplied with $[f_1f_2] = [fg]$. Indeed, the operator string first acts as $D\otimes \one$, and then as $L\otimes R$.

The second order Dyson MPO can be made by starting from the modified $\Tilde{H}^2$ MPO, identifying the levels containing only $1$s and $3_a$s, and rerouting the arrows ending on these levels while multiplying by the correct time-ordered integral.

The correct time-ordered integral is prescribed by the subscripts of the $3$s in the level. Level $(13_1)$ encodes a first order term and should be multiplied by $[f_1]$. Level $(3_23_1)$ encodes a second order term and should be multiplied by $[f_2f_1]$. Because $(1) \rightarrow (13_1)$ and $(1)\rightarrow(3_11)$ encode the same operator string, we additionally multiply it by $1/2$ to avoid double counting.

\section{Compression}
\label{sec:compression}

Just like the Taylor MPO, the Dyson MPO can be compressed to a much smaller bond dimension without compromising its accuracy. In a first \emph{exact} compression step, we identify equivalent columns. In a second \emph{approximate} compression step we further compress it by identifying similarities in the rows. The second compression technique is approximate but only introduces extra errors of an order higher than the Dyson MPO itself.

\subsection{Equivalent Columns}

There are a number of columns in the Dyson MPO that are completely equivalent. If we add together the corresponding rows\footnote{Rows labeled with the same level as the columns.} -- which encode the operators that follow the ones in the columns -- we can remove all but one of the equivalent columns. This therefore corresponds to an exact compression.

The basic observation underlying the compression steps is that columns whose labels are the same tuple after removing all 1's are compressible. The reason is that the $1$'s in the levels represent terms in the MPO that have not started yet. Moving the $1$'s through the column label will not change the operators that came before it. All levels that are the same tuple after removing all $1$'s have identical ``histories'' and can therefore be compressed.

We illustrate this by looking at the FSM representation of the third order Dyson MPO for the simple example of Sec.~\ref{sec:Dyson-Illustration}. In the third-order MPO, there will be three different levels labelled, e.g., by $(12_13_2)$, $(2_113_2)$ and $(2_13_2 1)$, but these all have the same history:
\begin{equation}
\includegraphics[page=1, width=\linewidth]{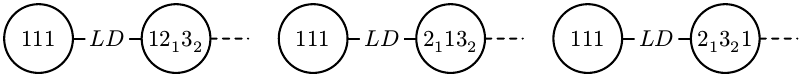}
\end{equation}
Therefore, these levels can be compressed to a single level labelled by $(2_12_3)$ :
\begin{equation} \label{eq:column_compressed}
\raisebox{-0.45\height}{\includegraphics[page=2, width=0.32\linewidth]{figures/compressionplots.pdf}} \;.
\end{equation}
This amounts to adding the rows labelled by $(2_113_2)$ and $(2_13_2 1)$ to the row labelled by $(12_13_2)$ (which are represented by the dotted lines in the above figures), and then removing levels $(2_113_2)$ and $(2_13_2 1)$ from the MPO entirely.\footnote{This can be implemented as a gauge transform in the MPO. Let $\mathcal{L}$ denote a set of virtual levels, such that for a given cut on a certain bond the left half of the MPO associated with levels $l \in \mathcal{L}$ evaluates to the same operator string. We can then insert an identity in the form of
\begin{equation}
I = \begin{bmatrix}
1 & -1 & -1 & \cdots & -1\\
  & 1 & 0 & \cdots & 0\\
  &   & 1 & \cdots & 0\\
  &   &   & \ddots & \vdots\\
  &   &   &   & 1
\end{bmatrix} \begin{bmatrix}
1 & 1 & 1 & \cdots & 1\\
  & 1 & 0 & \cdots & 0\\
  &   & 1 & \cdots & 0\\
  &   &   & \ddots & \vdots\\
  &   &   &    & 1
\end{bmatrix}
\end{equation}
and absorb the left upper triangular matrix into the tensor to the left of the cut. This makes all operator strings on the left half zero, except for the first one which is kept intact. Absorbing the right upper triangular matrix into the MPO tensor to the right of the cut has the effect of summing all rows associated with $\mathcal{L}$ into the first levels, whereas the other levels are unaffected but can be discarded because they are coupled to zero operator strings.} We denote the compressed level as \((2_1 3_2)\) for simplicity and we remember that the operators that follow this level correspond to the sum of the operators following $(12_1 3_2)$, $(2_1 13_2)$ and $(2_1 3_2 1)$.

Similarly, the levels labelled by $(12_13_1)$, $(2_113_1)$ and $(2_1 3_1 1)$ share the same history,
\begin{equation}
\includegraphics[page=3, width=\linewidth]{figures/compressionplots.pdf}
\end{equation}
and can therefore all be compressed to single level
\begin{equation}
\raisebox{-0.47\height}{\includegraphics[page=4, width=0.32\linewidth]{figures/compressionplots.pdf}} \;.
\end{equation}
in the MPO.

This procedure of compressing levels with the same history of operators can be generalized to any order in $N$ and for any form of the Hamiltonian.

\subsection{Approximate Row Compression}

In the exact column compression above, different levels that have the same history can be compressed because all operator strings to the left of the given cut are entirely equivalent. As a result, we have now a truncated MPO where the levels are labelled by compressed labels as in Eq.\eqref{eq:column_compressed}. Starting from this MPO representation, a similar procedure can now be carried out by considering the operator strings to the right of a given cut. Indeed, the operators that follow the row labels are often the same, but are multiplied with different coefficients. Take the following elements in the MPO for example:
\begin{align}
    &O[(2_1) , ~(3_1)] = [f_1]R^{(1)}\\
    &O[(2_13_1), ~(3_13_1)] = [f_1f_1]R^{(1)} \;.
\end{align}
Introducing the notation for an operator: \((2_1)\rightarrow(3_1) \equiv R^{(1)}\), we say that \((2_1)\) has coefficient \([f_1]\) along operator \((2_1)\rightarrow(3_1)\). In the same way the column-compressed level \((2_13_1)\) has coefficient \([f_1f_1]\) along \((2_1)\rightarrow(3_1)\). We can see thus that both levels $(2_1)$ and $(2_13_1)$ are followed by the same operator but with a different coefficient. For that reason, a compression is possible.

A fully systematic way of finding all the possible compressions of this form is given by the following procedure. The idea is to proceed iteratively through all the levels of the MPO and build up a set of \textit{kept levels} : a basis $\gamma_\text{kept}$ that spans the right-half operator space of the full MPO up to the required order. The way to proceed through all the MPO levels is to group them by increasing number of 2-indices ($n_2$) and 3-indices ($n_3$), and loop over this ordered set $(n_2,n_3)$ sequentially. In each iteration, all levels that are compatible with the group label $(n_2,n_3)$ are found and we ask whether their operator content is already captured by the kept levels accumulated so far. To answer this, we construct the representation of these compatible levels with respect to their combined operator content, which we call \(\gamma_{\text{compatible}}\). Then we express the kept levels in this basis, constructing \(\gamma_{\text{kept}}\). We then construct a projector onto the column space of $\gamma_\text{kept}$, and apply a rank-revealing QR decomposition to the residual $(\one - P_{\gamma_\text{kept}})\gamma_\text{compatible}$ to identify which compatible levels introduce genuinely new operators. Those that do are added to the kept set, the remainder are expressed as linear combinations of the kept levels by solving the corresponding linear system, and their columns in the MPO are added to the ones of the kept levels according to the linear combination and removed from the MPO. In this way, by looping over all groups of MPO levels we indeed build a full basis $\gamma_\text{kept}$ and express all MPO entries in terms of these basis elements with the appropriate coefficients.

The key observation that makes this step approximate rather than exact is that we only look at the operators following the to-be-compressed -- ``compatible" -- levels. The kept levels may have additional operators following them beyond those tracked in this basis. However, any extra terms thereby introduced carry prefactors of order strictly higher than $N$, and therefore do not affect the accuracy of the $N$th order Dyson MPO. Hence, the approximate nature of the row compression algorithm has as a consequence that the fully compressed $N$th order Dyson MPO no longer exactly captures all higher order terms made up of disjoint lower order terms. Nonetheless, the resulting MPO is still accurate up to $N$th order while having a significantly reduced bond dimension.

To further clarify this construction, we write out the full row compression algorithm for a 3rd order Dyson MPO in Appendix~\ref{app:rowcompression}. As remarked at the end of Section~\ref{sec:taylor}, the two compression algorithms explained here can also be applied to the Taylor MPO, leading to slightly lower bond dimensions than in the original formulation of Ref.~\onlinecite{VanDamme2024}, as summarized in Table~\ref{tab:compression} of Appendix~\ref{app:rowcompression}.

\section{Benchmarks}
\label{sec:benchmarks}

Let us now benchmark the accuracy of the Dyson-MPO construction by considering a few time-modulated Hamiltonians of spin-1/2 chains, both on a small finite system (where we can compare to quasi-exact time evolution schemes), and on a system directly in the thermodynamic limit (where we compare to standard time-evolution schemes with extremely small time steps).

\subsection{Finite system size}

First we show the accuracy of the Dyson MPO on a finite system for which we can perform quasi-exact computations. In order to demonstrate that the Dyson MPO of $N$th order provides a time-evolution operator that has an error of $\mathcal{O}(dt^N)$, we apply it to a well-chosen initial state and compare to the result of a full state-vector simulation performed with QuantumOptics.jl \cite{kramerQuantumOpticsjlJuliaFramework2018} using a 9th order Verner integration scheme from DifferentialEquations.jl \cite{rackauckasDifferentialEquationsjlPerformantFeatureRich2017} and a time step of $10^{-3}$.

The Hamiltonian used is a modulated transverse-field Ising model
\begin{equation}
    \label{eq:modulated-ising}
    H(t) = \sin(\omega t)\sum_{i = 1}^{L-1} \sigma^z_i\sigma^z_{i+1} + \cos(\omega t)\sum_{i=1}^L \sigma^x_i
\end{equation}
We take \(\omega = 2\pi\) and evolve the $\ket{0}^{\otimes L}$ product state for one period (\(t = 0\rightarrow 1\)) on a system of $L=8$ sites. The error measure \(\varepsilon\) is taken to be the trace distance:
\begin{equation} \label{eq:epsilon_finite}
    \varepsilon(\psi_A, \psi_B) = \sqrt{1 - \left|\braket{\psi_A |\psi_B}\right|^2}.
\end{equation}
The MPOs were applied to the MPS using a finite version of the algorithm of Ref.~\onlinecite{Vanhecke2021b}. We allowed the bond dimension to reach its maximal value, which equals $D=16$ for this system of eight sites. The only source of error is therefore given by the Dyson-MPO construction.

\begin{figure}
    \centering
    \includegraphics[width=\linewidth]{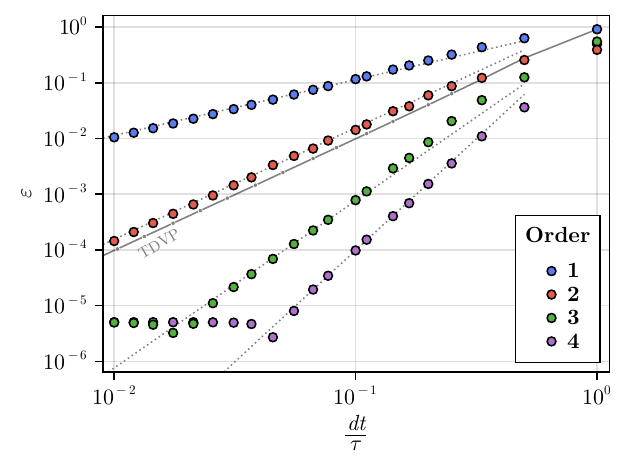}
    \caption{Benchmark for finite-system size. We plot the error measure $\varepsilon$ [Eq.~\eqref{eq:epsilon_finite}] of the state evolved with the Dyson-MPO of different orders with respect to a quasi-exact simulation. We evolve the state $\ket{0}^{\otimes L}$ over one period of the Hamiltonian of Eq.~\eqref{eq:modulated-ising}. The quasi-exact simulation was generated with QuantumOptics.jl as a function of the time step.}
    \label{fig:finite-scaling}
\end{figure}

The results are shown in Fig.~\ref{fig:finite-scaling}. The dotted lines are proportional to different powers of \(dt\) and show that the Dyson MPO of $N$th order provides a time-evolution operator that has an overall error scaling as \(\mathcal{O}(dt^N)\). The plateau of the third and fourth order errors can be attributed to the finite accuracy of the comparison wavefunction from QuantumOptics.jl. This is confirmed by comparing all results against the fourth-order DysonMPO-based result generated with the smallest time step, in which case the data continues on its straight line (see Appendix~\ref{sec:noplateau}).

\subsection{Infinite system size}

Next, we perform time evolution on an infinite spin-1/2 chain with the Heisenberg XXZ model featuring a time-modulated anisotropy
\begin{equation}
\label{eq:modulated-xxz}
H(t) = \sum_{\braket{ij}} \sigma^x_i \sigma^x_j + \sigma^y_i \sigma^y_j + \Delta(t) \sigma^z_i \sigma^z_j ,
\end{equation}
with
\begin{equation}
    \Delta(t) = 2 + \sin(\omega t) \;.
\end{equation}
We take $\omega = 2\pi$ and evolve the Néel product state for one period (\(t = 0\rightarrow 1\)). We compare the final state to a reference state obtained from a TDVP simulation with \(dt = 10^{-7}\) corresponding to dividing the period into a million time steps. The error measure \(\varepsilon\) is now a modified version of the trace distance:
\begin{equation} \label{eq:error-infinite}
    \varepsilon(\psi_\text{A}, \psi_\text{B}) = \sqrt{1-\left|\lambda^{\text{A}}_\text{B}\right|^2},
\end{equation}
where \(\lambda^{\text{A}}_\text{B}\) is the leading eigenvalue of the mixed transfer matrix of the MPS parametrized by the tensors \(A\) and \(B\) [see Eq.~\eqref{eq:mixedtransfermatrix}]. This is a proper definition of the distance between states in the thermodynamic limit that avoids the orthogonality catastrophe \cite{vanderstraeten2019}. We use a variational algorithm for implementing the MPO-MPS application \cite{Vanhecke2021b}, with a maximal allowed bond dimension of $D=64$ making sure that all results have converged.

\begin{figure}
    \centering
    \includegraphics[width=1\linewidth]{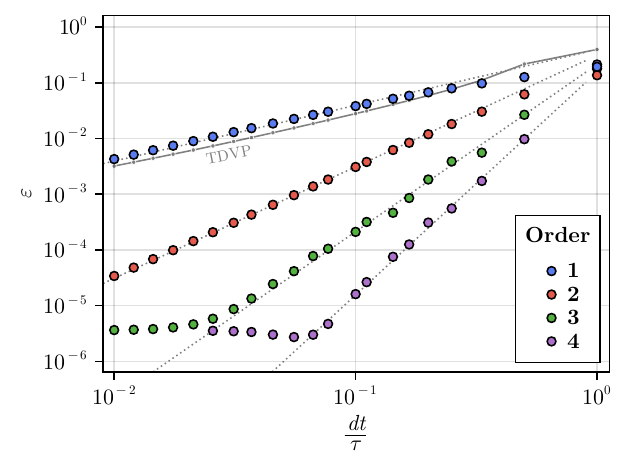}
    \caption{Benchmark for infinite-size system. We plot the error measure $\varepsilon$ [Eq.~\eqref{eq:error-infinite}] of the state evolved with the Dyson-MPO of different orders, with respect to a TDVP simulation. We evolve the Néel state over one period of the Hamiltonian of Eq.~\eqref{eq:modulated-xxz}. The infinite-size TDVP simulation is performed with a time step of $dt=10^{-7}$.}
    \label{fig:infinite-scaling}
\end{figure}

The results are shown in Fig.~\ref{fig:infinite-scaling}. The dotted lines are again proportional to different powers of \(dt\). Once more, they show that the Dyson MPO of $N$th order provides a time-evolution operator that has an overall error scaling as \(\mathcal{O}(dt^N)\). Here too, the plateau in the error for the third and fourth order can be attributed to the finite accuracy of the comparison wavefunction. In Appendix~\ref{sec:noplateau} we use our most accurate Dyson MPO generated wavefunction as a reference to confirm this.

\subsection{Runtime advantage}

\begin{figure}
    \centering
    \includegraphics[width=1\linewidth]{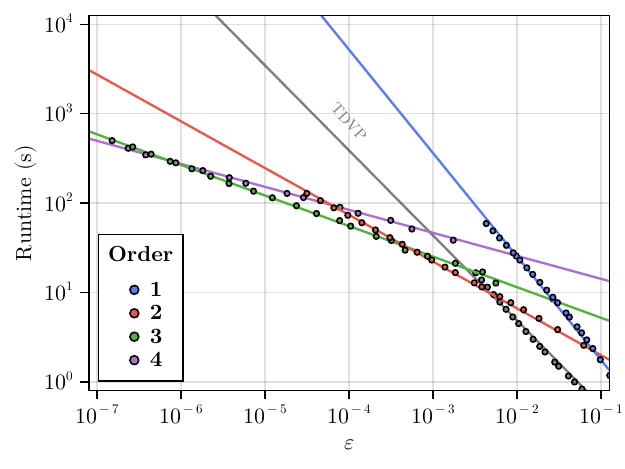}
    \caption{Runtime comparison. We compare the total runtime for the results of Fig.~\ref{fig:infinite-scaling} as a function of the achieved accuracy.}
    \label{fig:runtime}
\end{figure}

While the improved accuracy for a given time step already shows a clear advantage over a conventional method like TDVP, accuracy alone does not fully capture the practical performance of a time-evolution method. In practice, the relevant figure of merit is the computational cost required to reach a given target accuracy. In Fig.~\ref{fig:runtime}, we therefore compare the runtime as a function of the achieved error for the different orders, again for the same benchmark as in the previous section. We now compare to the most accurate Dyson MPO generated wavefunction to eliminate the finite-accuracy effect that led to the plateaus in Fig.~\ref{fig:finite-scaling} and Fig.~\ref{fig:infinite-scaling}.

Fig.~\ref{fig:runtime} shows that the enhanced accuracy of our approach translates directly into a more favourable runtime-accuracy trade-off, yielding higher accuracy at equal computational cost, or equivalently, a reduced runtime for a fixed accuracy threshold.

These simulations were all performed on the same machine. Each simulation was allowed to use one Julia thread and four BLAS threads. The runtime was taken to be \(N_{\text{steps}} \times \Delta t_{\text{av}}\) where \(\Delta t_{\text{av}}\) was calculated as the average time it takes for one time step, averaged over a thousand steps.

\section{Conclusion and Outlook}
\label{sec:outlook}

We have introduced an MPO encoding of the Dyson series for (quasi) one-dimensional lattice systems, which is the time-evolution operator for systems with time-dependent Hamiltonians. We can systematically increase the order of truncation in the Dyson series, reducing the scaling of the error with respect to the time step. Our proposal builds upon previous work that implements the exponential of Hamiltonians as an MPO and inherits its beneficial properties: (i) it is an extensive operator that can be used directly in the thermodynamic limit, and (ii) it can handle both local and quasi local operators, making it suitable for systems with long-range interactions.

Several directions remain open for future work. The present construction opens the door to more efficient schemes for handling time-dependent time-evolution problems, for instance by combining the Dyson MPO with adaptive time-stepping strategies by exploiting structure in the time-dependence of the Hamiltonian to further reduce computational cost. A natural extension is the application of the Dyson MPO to periodically driven systems, where one could simulate the dynamics stroboscopically, evaluating the state only at integer multiples of the driving period. This would allow one to extract the long-time behavior of Floquet systems without resolving the micromotion. Related to this, it would be interesting to study which additional information can be extracted from the MPO representation of a full driving cycle.

In a parallel development, we have also shown that it is possible to systematically formulate the Magnus expansion in the format of a first-degree MPO. Because this work was aimed at simulating time evolution, this idea was set aside for the Dyson series, which is a more natural and efficient formulation of the time-evolution operator for time-dependent systems. However, being able to formulate the Magnus expansion as a first-degree MPO, directly in the thermodynamic limit and for arbitrarily long-range interactions, is a powerful tool which could prove useful in the study of Floquet dynamics of driven systems. In particular, given an efficient MPO representation of the Floquet Hamiltonian up to arbitrary order, one could study the ground state and excitation spectrum of the effective Hamiltonian using conventional MPS methods.

We believe that our MPO constructions offer a natural framework for benchmarking and optimizing quantum circuits in digital quantum simulators. Since the Dyson MPO provides a systematic, order-by-order approximation to the time-evolution operator, it could be used to assess the accuracy of a given circuit ansatz or to guide the compilation of time-evolution circuits for time-dependent Hamiltonians.

It would also be worthwhile to explore whether our Dyson and Magnus encodings can be generalized to higher-dimensional systems through the use of projected entangled-pair operators (PEPOs), the two-dimensional analogue of MPOs. Such a generalization would significantly broaden the applicability of the method to a wider class of strongly correlated quantum systems.

Finally, we can consider time-dependent perturbation theory of the form
\begin{equation}
    H(t) = H_0(t) + V_I(t),
\end{equation}
such that the time-evolution operator in the interaction picture can be represented in the form of a Dyson series. We can imagine a situation where the time evolution according to $H_0(t)$ can be very well represented by a tensor network (\textit{e.g.}~a sum of strictly one-site operators), and we want to approximate the influence of an additional small time-dependent term $V_I(t)$. Our scheme for the Dyson series can be applied to this situation as well.

\begin{acknowledgments}
V.V. thanks Lukas Devos and Lander Burgelman for a previous collaboration that inspired this project. We also acknowledge inspiring discussions with Frank Verstraete and Ignacio Cirac. V.V. was supported by the Research Foundation Flanders (FWO) under doctoral fellowship No.~1196525N. I.P.M.~acknowledges funding from the National Science and Technology Council (NSTC) Grant No.~113-2112-M-007-021-MY2.
\end{acknowledgments}

\bibliography{dyson}

\clearpage\newpage
\onecolumngrid

\appendix

\section{Approximate row compression}
\label{app:rowcompression}

In this appendix, we illustrate the row compression algorithm concretely for a 3rd order Dyson MPO with two driving channels: $f_1(t)$ driving a two-site operator and $f_2(t)$ driving an on-site operator, i.e.
\begin{equation}
    H(t) = f_1(t)H^{(1)} + f_2(t)H^{(2)}
\end{equation}
where \(H^{(1)}\) is a sum of two-body operators (i.e.\ \(\sum_i L_i\otimes R_{i+1}\)) and \(H^{(2)}\) is a sum of one-site operators (i.e.\ \(\sum_i D_i\)). The modified Hamiltonian $\tilde{H}$ has levels $(1), (2_1), (3_1)$ and $(3_2)$, with $(2_1)$ only transitioning to $(3_1)$:
\begin{equation}
\tilde{H}(t) \sim \begin{pmatrix} \one & L & 0 & f_2(t) D \\ & 0 & f_1(t) R & 0 \\ & & \one & 0 \\ & & & \one
    \end{pmatrix}
\end{equation}

The algorithm iterates over all levels in the MPO, grouped by the number of 2-indices ($n_2$) and 3-indices ($n_3$) they contain, looping as $n_2 = \{1, 2, \ldots, N\}$ and $n_3 = \{0, 1, \ldots, N - n_2\}$. At each step we maintain a set of \textit{kept levels} that collectively span the operator space of the right half of the MPO up to the required order. These operators are found as ways of transitioning from the $2_i$s and the $N-n_2 - n_3$ $1$s in the current level $l$ (across an arbitrary number of sites) to a level $l'$ consisting of all $3_j$s and $1$s, where all $1$ labels are implicit because of the earlier steps (including the column compression) in the MPO construction. The different operators will be schematically denoted as $l \to l'$. For each new group of compatible levels ---levels that are compatible with \(n_2\) and \(n_3\)--- we check whether they introduce any operator content not already captured by the kept set, add those that do, and eliminate the rest by expressing them as linear combinations of the kept levels. For a given value of $n_2$, we start with $n_3=0$ as this contains the most operators (associated with adding implicit $1$s), and subsequently consider cases with increasing $n_3$.

The tables below display, for each iteration, the coefficient matrix $\gamma$ whose columns correspond to compatible or kept levels and whose rows correspond to operator basis elements, while the entries are the time-ordered integral prefactors associated with each level-operator pair. We can interpret $\gamma^\mathrm{T}$ as representing the full right-half MPO after cutting one bond. We aim to decompose $\gamma$ via a selection of a subset $S$ of its columns, corresponding to the ``kept'' levels, as $\gamma = \gamma_{:,S} X = \gamma I_{:,S} X$, where $I$ is the identity matrix and $X$ contains the linear expansion coefficients of all levels with respect to the kept levels. This decomposition implies that we can insert $X^{\mathrm{T}} (I_{:,S})^\mathrm{T}$ on every bond in the MPO, which ---after absorbing these two factors to the left and right MPO tensor respectively--- amounts to only keeping the rows in $S$ and applying the linear combination encoded in $X^{\mathrm{T}}$ to the columns.

We always keep the ``\((1)\)" level, and therefore omit it from the following discussion. Proceeding, we show this looping over two- and three-levels, each iteration has the same structure: we present the current kept set, the candidate levels, and determine which to absorb by looking at their representation with respect to the operator basis of operators following the compatible levels.
\begin{flalign*}
& & \includegraphics[page=8, width=0.97\linewidth]{figures/compressionplots.pdf} &
\end{flalign*}
As there are no kept levels yet to express the compatible level in, we keep it.
\begin{flalign*}
& & \includegraphics[page=9, width=0.97\linewidth]{figures/compressionplots.pdf} &
\end{flalign*}
To figure out which of the compatible levels we need to keep in order to span their complete operator basis we perform two steps:
\begin{itemize}
\item[(i)] Construct a projector onto the column space of the kept levels' representation \(\gamma_{\text{kept}}\). Such a projector can be constructed as \(P_{\gamma} = \gamma\cdot(\gamma^{\dagger}\gamma)^{-1}\cdot \gamma^{\dagger} \).
\item[(ii)] Find a minimal subset of compatible levels that span the rest of the operator space by performing a rank-revealing QR decomposition on \((\one - P_{\gamma_{\text{kept}}})\cdot \gamma_{\text{compatible}}\).
\end{itemize}
The rank-revealing QR decomposition allows for a tolerance which may help in further compressing the Dyson MPO if, at some point of the time evolution, some integrals turn out to be vanishing. We use a rank-revealing QR decomposition because it employs a pivoting strategy which allows us to identify which levels we need to add to our kept basis to span the full operator space, without changing the levels themselves. After identifying the compatible levels that span the extra part of the operator space (if there are any), we add them to the kept levels, and express all levels as a linear combination $X$ of the kept levels. The columns of $X$ corresponding to the kept levels themselves are of course trivial (having a single one and all zeros), whereas the columns associated with the other levels, collectively denoted as $x$, are obtained by solving the linear system \(\gamma_{\text{kept}}x = \gamma_{\text{rest}}\), with \(\gamma_{\text{rest}}\) holding the column vectors of the non-kept compatible levels.

In this case it turns out that we need to keep two levels, \((2_1 3_1)\) and \((2_1 3_2)\) and that we can express the other compatible levels as:
\begin{align}
    \label{eq:compression-explicitlinco}
    (3_1 2_1) &= [f_1] \cdot (2_1) - (2_1 3_1)\\
    (3_2 2_1) &= [f_2] \cdot (2_1) - (2_1 3_2),
\end{align}
\noindent because of the identities:
\begin{align}
    &[a][b] = [ab] + [ba] \\
    &[ab][c] = [abc] + [acb] + [cab].
\end{align}

By only looking at the basis of operators that follow the compatible levels up to order $N$, we have ignored some of the operators that follow \((2_1)\). By compressing the Dyson MPO we have, however, added new terms to the operator corresponding to these ``ignored" operators. An operator that follows \((2_1)\) (but not \((2_1 3_1)\)) is \((112_1) \rightarrow (3_13_13_1)\). Because of Eq.~\eqref{eq:compression-explicitlinco} we have introduced a term \([f_1][f_1f_1f_1]\cdot((112_1) \rightarrow (3_13_13_1))\), which is of fourth order. It is the fact that we only look at the operator basis of the compatible levels, and not at the full operator basis of both the kept and compatible levels, that manifestly allows us to only perform a compression step that is only valid up to $N$th order. All extra terms we add by ignoring part of the kept level's operator basis are of an order higher than the Dyson MPO itself, which makes the row compression correct up to the order of the Dyson MPO.

\noindent We continue the loop with the next iteration:
\begin{flalign*}
 & & \includegraphics[page=10, width=0.97\linewidth]{figures/compressionplots.pdf} &
\end{flalign*}
In this case, all compatible levels can be removed, because their columns lie in the span of the kept columns in the truncated right-operator basis. We then proceed to the next iteration:
\begin{flalign*}
 & & \includegraphics[page=11, width=0.97\linewidth]{figures/compressionplots.pdf}&
\end{flalign*}
As the level \((2_12_1)\) introduces an operator which is not yet present in the operator basis of the kept levels, we have to keep it. We then consider the next group:
\begin{flalign*}
 & & \includegraphics[page=12, width=0.97\linewidth]{figures/compressionplots.pdf}&
\end{flalign*}
Again, all compatible levels can be removed. The final iteration is:
\begin{flalign*}
 & & \includegraphics[page=13, width=0.97\linewidth]{figures/compressionplots.pdf}&
\end{flalign*}
As the level \((2_12_12_1)\) introduces an operator which is not yet present in the operator basis of the kept levels, we have to keep it.

This concludes the construction of the compressed Dyson MPO. In total we have kept the levels $(1)$, $(2_1)$, $(2_13_1)$, $(2_13_2)$, $(2_12_1)$ and $(2_12_12_1)$, which brings the total bond dimension of this 3rd order Dyson MPO to \(1 + 3\chi + \chi^2 + \chi^3\), where \(\chi\) is the dimension of the \((2_1)\) level. In general the bond dimension will depend on the amount of driving functions in the Hamiltonian.

Note that in Ref.~\onlinecite{VanDamme2024}, a slightly different compression approach was used. We can use the compression approach introduced in this section in the time-independent case by replacing all time-ordered integrals \([f_1f_2f_3\dots f_n]\) by \(\frac{\tau^n}{n!}\). It turns out that our construction yields a significant improvement, with the bond dimensions it achieves for the time-independent (Taylor) case being summarized in the Table~\ref{tab:compression}.

\begin{table}[t]
\begin{tabular}{|c|l|}
\hline
Order & Taylor MPO Bond Dimension                         \\ \hline
1     & $1 + \chi$                                        \\ \hline
2     & $1+ \chi + \chi^2$                                \\ \hline
3     & $1 + 2 \chi +  \chi^2 + \chi^3$                   \\ \hline
4     & $1 + 2 \chi + 3\chi^2 + \chi^3 + \chi^4$          \\ \hline
5     & $1 + 3\chi + 3\chi^2 + 4\chi^3 + \chi^4 + \chi^5$ \\ \hline
6     & $1 + 3\chi + 6\chi^2 + 4\chi^3 + 5\chi^4 + \chi^5 + \chi^6$ \\ \hline
\end{tabular}
\caption{Bond dimension of the Taylor MPO of order $N=1,2,\ldots,6$ for a time-independent Hamiltonian after applying the row compression algorithm discussed in Section~\ref{sec:compression} of the current manuscript.}
\label{tab:compression}
\end{table}

\section{Derivatives of time-dependent MPS or MPOs}\label{app:MPSderivatives}
Consider a (finite or infinite) MPS (or MPO) $\ket{\Psi(t)}$ with site tensors $\{A_i(t)\}$ and boundary vectors $v_L$ and $v_R$ (the precise value of which does not matter in the case of an injective MPS in the thermodynamic limit). It readily follows that the first derivative $\frac{d\ }{d t}\ket{\Psi(t)}$ is given by an MPS with site tensors and boundary vectors given by
\begin{align}
\tilde{A}_i &= \begin{bmatrix} A_i(t) & \dot{A}_i(t) \\ 0 & A_i(t)\end{bmatrix},&\tilde{v}_L &= \begin{bmatrix} v_L & 0 \end{bmatrix}, & \tilde{v}_R &= \begin{bmatrix} 0 \\ v_R \end{bmatrix}.
\end{align}
Applying this to the first-order Taylor MPO with tensor $O(\tau) = \begin{bmatrix} \one + \tau D & L \\ \tau R & A \end{bmatrix}$ and arbitrary boundary vectors that we can choose as $v_L = \begin{bmatrix} 1 & 0\end{bmatrix}$ and $v_R = \begin{bmatrix} 1 \\ 0 \end{bmatrix}$ and evaluate the result at $\tau =0 $, we obtain an MPO with tensors
\begin{align}
\tilde{O} &= \begin{bmatrix} \one & L & D & 0 \\ 0 & A & R & 0 \\ 0 & 0 & \one & L \\ 0 & 0 & 0 & A \end{bmatrix},&\tilde{v}_L &= \begin{bmatrix} 1 & 0 & 0 & 0 \end{bmatrix}, & \tilde{v}_R &= \begin{bmatrix} 0 & 0 & 1 & 0 \end{bmatrix}^T.
\end{align}
It can easily be seen that the fourth row and column can never be reached, and can thus be discarded. This is exactly the MPO representation of the Hamiltonian, thus showing that the first order Taylor MPO indeed satisfies the Schr\"odinger equation for the evolution operator at $\tau = 0$.

This strategy can be generalized to higher orders. The second derivative of an MPS with general tensor $\{A_i(t)\}$ is then given by an MPS with tensors
\begin{align}
\tilde{\tilde{A}}_i &= \begin{bmatrix} A_i(t) & \dot{A}_i(t)  & \dot{A}_i(t) & \ddot{A}_i(t)\\ 0 & A_i(t) & 0 & \dot{A}_i(t) \\ 0 & 0 & A_i(t) & \dot{A}_i(t) \\ 0 & 0 & 0 & A_i(t)\end{bmatrix},&\tilde{\tilde{v}}_L &= \begin{bmatrix} v_L & 0 & 0 & 0\end{bmatrix}, & \tilde{\tilde{v}}_R &= \begin{bmatrix} 0 \\ 0 \\ 0 \\ v_R \end{bmatrix}.
\end{align}
Observing that, after applying the gauge transform
\begin{equation}
G = \begin{bmatrix} 1 & 0 & 0 & 0 \\ 0 & \frac{1}{\sqrt{2}} & \frac{1}{\sqrt{2}} & 0 \\ 0 & \frac{1}{\sqrt{2}} & -\frac{1}{\sqrt{2}} & 0 \\ 0 & 0 & 0 & 1 \end{bmatrix}
\end{equation}
row and column 3 cannot be reached and can thus be discarded, and extracting a factor $\frac{1}{2!}$, leads to the compactified representation for the tensors of an MPS representation of $\frac{1}{2!} \frac{d^2 \ket{\Psi(t)}}{d t^2}$
\begin{align}
\tilde{\tilde{A}}_i &= \begin{bmatrix} A_i(t) &  \dot{A}_i(t)  & \frac{1}{2!}\ddot{A}_i(t)\\ 0 & A_i(t) &  \dot{A}_i(t) \\ 0 & 0 & A_i(t) \end{bmatrix},&\tilde{\tilde{v}}_L &= \begin{bmatrix} v_L & 0 & 0\end{bmatrix}, & \tilde{\tilde{v}}_R &= \begin{bmatrix} 0 \\ 0 \\ v_R \end{bmatrix}.
\end{align}
The structure can be extended to higher order and generalises a well-known result for the derivative of a (meromorphic) function of a one-parameter family of matrices $A(t)$
\begin{align}
\frac{1}{p!}\frac{d^p\ }{d t^p} f(A(t)) = \begin{bmatrix} \one & 0 & \cdots & 0 & 0 \end{bmatrix}
f\left(\begin{bmatrix} A & A^{(1)} & \cdots & \frac{1}{(p-1)!} A^{(p-1)} &\frac{1}{p!} A^{(p)} \\
0 & A & \cdots & \frac{1}{(p-2)!} A^{(p-2)} & \frac{1}{(p-1)!} A^{(p-1)}\\
\vdots & \vdots & \ddots & \vdots & \vdots \\
0 & 0 & \cdots & A & A^{(1)}\\
0 & 0 & \cdots & 0 & A
\end{bmatrix}\right) \begin{bmatrix} 0 \\ 0 \\ \vdots \\ 0 \\ \one \end{bmatrix}
\end{align}
with $A^{(k)} = d^k A(t) / d t^k$. This result thus remains to hold when the entries of $A$ are not scalars, but single-site states or operators, which are then composed using tensor products upon multiplication of the matrix.

\section{Time-ordered integrals using the Quantics approach}
\label{sec:timeorderedintegrals}
One can significantly speed up the calculation of time-ordered integrals by making use of a quantics representation of the driving functions. Many functions, such as \(\sin(x)\), \(\cos(x)\), and \(e^x\) permit efficient encodings in terms of a quantic tensor train \cite{waintal2026}. Others can be learned through tensor cross interpolation.

Integration over a high dimensional integration domain using conventional techniques like Gaussian quadrature become increasingly intractable the higher the dimension. In the quantic tensor train representation many operations like derivatives and integrals are surprisingly inexpensive.

Therefore, we use the quantics representation to perform the time-ordered integrals needed for the construction of the Dyson MPO.
The code for which can be found on github at \href{https://github.com/VictorVanthilt/QuanticsTT.jl}{github.com/VictorVanthilt/QuanticsTT.jl}.

\subsection{The quantic tensor train representation of functions}
To encode a function \(f(x)\) on the domain \([0, 1)\), we divide the domain in \(2^N\) equally spaced points \(x_n = \frac{n}{2^N}\) where \(n\in \{0, ..., 2^N - 1\}\). We denote \(f_n \equiv f(x_n)\). We can express \(n\) in terms of its binary representation \(n_{N-1}\dots n_1n_0\) as \(n = \sum_{\alpha = 0}^{N-1} n_{\alpha} 2^{\alpha}\) where \(n_{\alpha} \in \{0, 1\} \; \forall \alpha\).

There are several functions that allow for an efficient encoding as a tensor train. Take the exponential function for example:
\begin{equation}
    e^{x_n} = \exp\left(\sum_{\alpha=0}^{N-1} n_\alpha 2^{\alpha-N} \right),
\end{equation}
which can be written as a product over \(\alpha\) as:
\begin{equation}
    e^{x_n} = \prod_{\alpha=0}^{N-1}\exp\left(n_{\alpha}2^{\alpha - N}\right)
\end{equation}
The quantics tensor train representation of \(e^x\) is thus of bond dimension 1. The entries of the tensors are:
\begin{equation}
    \includegraphics[page=2]{figures/quantics.pdf} \;.
\end{equation}
The sites are now associated with the different bits of the function argument.  Note that we have chosen the least significant bit to be on the first site, and the most significant on the last.

To evaluate the function at a point \(x_n\), one fills in the binary representation of \(x_n \cdot 2^N\) on the ``physical" legs of the tensor train and evaluates the network.
\begin{equation}
    \includegraphics[page=3]{figures/quantics.pdf}
\end{equation}

As \(\sin(x) = \frac{e^{ix} - e^{-ix}}{2i}\), its quantics representation is a bond dimension \(2\) tensor train and manipulating it will be computationally cheap.

\subsection{Integration using the quantics representation}

One can perform indefinite integrals on quantic tensor trains by multiplying it by a rank 2 MPO.
The main idea is that indefinite integrals can be performed as:
\begin{equation}
    F(y_n) = \int_0^{y_n}f(x) dx = \sum_{m < n} f(x_m) \Delta x = \sum \Theta(n-m) f_m \Delta x
\end{equation}
We therefore need an implementation of the Heaviside function \(\Theta(n-m)\) as an MPO. To determine if an integer \(n\) is larger than \(m\), we can compare their binary representations starting from the most significant bit. If \(n_{2^N-1} > m_{2^N-1}\) we know that, regardless of the other bits, \(n>m\). If the bits are equal, the comparison is undecided, and we move left to the next significant bit. For the value of the indefinite integral, the bits of $x$ to the left of the rightmost bit where $0 = x_k < y_k =1$ can take any value, amounting to summing up all those contributions $m < n$. This simple algorithm can be encoded in an MPO of bond dimension 2, where the rightmost bit $0 = x_k < y_k =1$ instigates a transition between virtual level $0$ on its right (forcing all $x_{j} = y_{j}$ for $j>k$) and virtual level $1$ on its left (allowing any possible value of $x_j$ and $y_j$ for $j<k$). The MPO tensor is thus given by
\begin{equation}
\raisebox{-0.45\height}{\includegraphics[page=4]{figures/quantics.pdf}} = \delta_{a,1}\delta_{b,1}
+ \delta_{x,y}\delta_{b,0}\delta_{a, 0}  + (\Delta x) \delta_{y, 1}\delta_{x,0}\delta_{a,1} \delta_{b,0}
\end{equation}
and the resulting MPO should be terminated with $a=1$ at its leftmost virtual index and $b=0$ at its rightmost index.

The pointwise product of two functions $(h(x) = f(x) \cdot g(x))$ can be obtained with the simple MPO tensor:
\begin{equation}
    \raisebox{-0.45\height}{\includegraphics[page=5]{figures/quantics.pdf}} = \delta_{x,y,z} \;.
\end{equation}
These two elements can now be combined in an elementary building block of a time-ordered integral:
\definecolor{jblue}{RGB}{64, 99, 216}
\begin{equation}
    f(t)\cdot\int_0^t\textcolor{jblue}{g(}t_1\textcolor{jblue}{)}dt_1 = \raisebox{-0.45\height}{\includegraphics[width=0.35\linewidth,page=1]{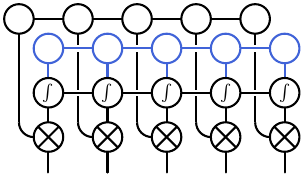}}
\end{equation}
Combining multiple steps of this operation allows for evaluating any time-ordered integral.

\section{Additional Figures}
\label{sec:noplateau}
\begin{figure*}
    \centering
    \includegraphics[width=0.49\linewidth]{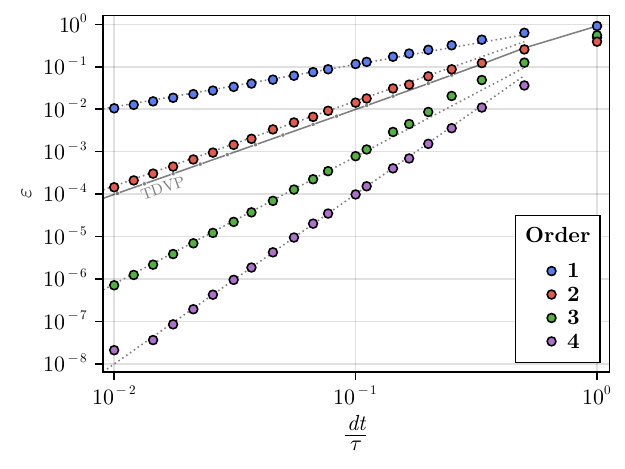}
    \includegraphics[width=0.49\linewidth]{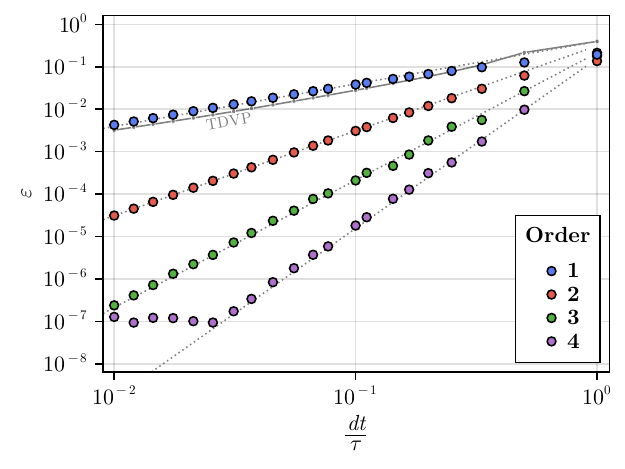}
    \caption{Finite-size benchmark (left) and infinite-size benchmark (right) using the most accurate wavefunction generated with the 4th order Dyson MPO with the smallest timestep as reference.}
\end{figure*}

\end{document}